# Bi-domain state in the exchange bias system FeF$_2$/Ni

O. Petracic*,[1,2], Zhi-Pan Li[1], Igor V. Roshchin[1], M. Viret[1,3], R. Morales[1,4], X. Batlle[1,5], and Ivan K. Schuller[1]

[1]Department of Physics, University of California – San Diego, La Jolla, CA, 92093-0319, USA

[2]Angewandte Physik, Universität Duisburg-Essen, 47048 Duisburg, Germany

[3]CEA Saclay, 91191 Gif sur Yvette cedex, France

[4]Departamento de Fisica, Universidad de Oviedo, Oviedo 33007, Spain

[5]Departament de Física Fonamental, Universitat de Barcelona, 08028 Barcelona, Catalonia, Spain

**Abstract** − Independently exchange biased subsystems can coexist in FeF$_2$/Ni bilayers after various field-cooling protocols. We find double hysteresis loops for intermediate cooling fields, while for small or large cooling fields a negatively or positively shifted single loop, respectively, are encountered. Both the subloops and the single loops have the same absolute value of the exchange bias field, $\mu_0|H_E| \approx 0.09$ T. This suggests that the antiferromagnet breaks into two magnetic subsystems with opposite signs but equal magnitude of bias acting on the ferromagnet. In this case the ferromagnet does not experience an average bias from the antiferromagnet but rather two independent subsystems ('bi-domain' state). This idea is confirmed by micromagnetic simulations including the effect of the antiferromagnet. We also present experiments, where thermally activated motion of these antiferromagnetic 'domain' boundaries can be achieved.

Antiferromagnetic/ferromagnetic (AF/FM) bilayers can show the so-called exchange bias (EB) effect, which manifests as a shift of the hysteresis loop along the field axis [1−4]. There is a consensus, that due to the exchange interaction across the interface between AF and FM, the FM experiences an unidirectional anisotropy, which leads to the loop shift. However, microscopic interpretations and theoretical models still diverge.

Several models assume the existence of domain walls parallel to the interface in either the FM or AF or in both [4−7]. In this case EB arises from a spring like winding and unwinding of the domains. In several other models the AF remains essentially frozen throughout the hysteresis cycle, while uncompensated AF interfacial spins give rise to the EB shift. In this case two problems are encountered: Using microscopically reasonable values for the parameters the calculated EB field is strongly overestimated. Second, this model predicts zero EB for perfectly compensated AF interfaces, while in experiment even large EB fields are observed [8]. It turns out, that these discrepancies can be resolved assuming correct values for the AF interfacial moment. Recent experiments show the existence of frozen AF moments at the interface that constitute about 5% of the total number of AF magnetic moments [9, 10]. Several ideas were proposed in order to explain the occurrence of those locked moments, i.e. dislocation induced domain walls in the AF of fractal shape [11, 12], a stress-induced piezomagnetic moment in the AF [13] or uncompensated AF spins at the interface due to the topology of AF grain sizes [14].

In this report we present the AF/FM bilayer system, $FeF_2$/Ni, which has the characteristic to form sufficiently large domains in the AF. Furthermore, the EB shift is exceptionally large compared to the coercive field. This leads to the possibility to observe very clearly the case, where the EB effect is not averaged over the FM/AF interface, but where the system breaks into two oppositely biased subsystems [15].

A $FeF_2$(83 nm)/Ni(17 nm)/Al(6 nm) multilayer was grown on a single crystalline $MgF_2$(110) substrate by e-beam evaporation. Prior to deposition the substrate was heated to 500°C first in vacuum for 1 hour and then in an oxygen atmosphere ($p = 1 \cdot 10^{-4}$ Torr) for additional 30 min to burn off the hydrocarbons from the surface. The $FeF_2$ layer ($T_N$ = 78 K) was deposited at a temperature of 300°C at a rate of 0.05nm/s and the Ni layer at 150°C at the same rate. As a protection against oxidation an Al layer was deposited finally at 150°C at a rate of 0.1nm/s. The base pressure was lower than $2 \cdot 10^{-7}$ Torr. From X-ray diffraction measurements one can identify, that $FeF_2$ grows epitaxially in the (110) orientation, whereas the Ni is polycrystalline. $FeF_2$(110) has compensated interfacial spins with the easy axis lying in-plane along the [001] direction [8]. Magnetization measurements were performed using a commercial SQUID magnetometer (Quantum Design). The sample was mounted with the AF easy axis parallel to the applied field, i.e. **H** || [001]. This axis was found to be also the easy axis of the FM. Figure 1 shows the magnetization, $M$ vs. $H$, measured by SQUID magnetometry at $T$ = 10 K after field cooling (FC) from 150 K in $\mu_0 H_{FC}$ = 0.05 T (solid squares), 0.075 T (open triangles), 0.1 T (solid diamonds), 0.125 T (open squares) and 0.2 T (solid circles). For intermediate fields, 0.075 T $\leq \mu_0 H_{FC} \leq$ 0.125 T, one finds a double hysteresis loop, where the exchange bias and coercive fields of both subloops are almost identical, $|\mu_0 H_E| \approx$ 0.09 T and $\mu_0 H_c \approx$ 0.006 T, respectively, as determined from the inflection points. Upon application of a smaller, $\mu_0 H_{FC} \leq$ 0.05 T, or higher cooling field, $\mu_0 H_{FC} \geq$ 0.2 T, only a single loop with negative or positive EB shift, respectively, is observed. For lower or higher cooling fields (measured up to 7 T), no change is observed. One finds virtually no difference of the EB fields between the subloops and the single loops. The dependence of $H_E$ as a function of the FC field is presented in the inset. One finds a well-defined crossover region in which double hysteresis loops (DHL) appear, while the EB field varies only by 1% for the left hand side subloops and 7% for the right hand side subloops.

This hints toward a symmetrical biasing effect. Here we propose the existence of two subsystems (briefly called 'bi-domain') in the AF, where the pinned AF moments are oppositely oriented [13, 15–19]. Presently there is no hint, whether this subsystems are merely 'regions' consisting of a number of AF domains or whether they are real domains in the $FeF_2$. We further assume that the AF-FM exchange coupling is antiferromagnetic [8]. The consequence is, that each FM subsystem will experience either a negatively or positively shifted magnetization reversal corresponding to the AF subsystem. In our case the magnetization of the FM remains saturated during the FC procedure. This differs from several studies, where a double loop can only be found after demagnetizing the FM [16, 19].

We assume in our sample the mechanism similar to that found in other EB systems, where a transition from a negatively to a positively shifted loop is found depending on the magnitude of $H_{FC}$ [20]. This can qualitatively be understood in a simple picture, where only the uncompensated pinned AF moments at the AF/FM interface are considered. Then, for small $H_{FC}$ the AF interfacial moments are oriented completely by the AF-FM coupling energy leading to only one single subsystem in the AF which in turn yields only one negatively biased loop. Analogously, for large cooling fields the orientation of the AF interfacial moments is dominated by the applied field, which yields a single positively shifted loop. For intermediate cooling fields a crossover region is found, where both cases coexist. It is remarkable, that the EB value does not change significantly (see inset of Fig. 1). This is in contrast to the usually found behavior, where a gradual crossover from negative to positive EB is encountered [20]. Here we find a step-like behavior with a field region, where double loops appear. The difference must be due to the fact, that in our samples the FM does not experience an average AF moment $\langle S_{AF} \rangle$ [21] but rather two independent AF/FM subsystems coexist [15]. The net magnetization of the system is then simply a superposition of the two subsystems. In order to explain this bifurcation of the biasing directions, one needs either a variation of the coupling strength, $J_{AF-FM}$, over the AF-FM interface or a third energy term of the form $J_m(\mathbf{r}) \approx \pm J_m^0$. This additional term could be supplied by piezomagnetic or more generally magneto-elastic energies.

One should note, that no spin-flop scenario e.g. like in the single-crystal FeF$_2$/Fe system [22] occurs. This is evidenced from very rounded loops on a similar sample at both $T$ larger and smaller than $T_N$ measured along the hard axis. Interestingly the Ni layer shows even at 150 K > $T_N$ an easy axis along [001] (data not shown), although the X-ray diffraction data indicate a polycrystalline Ni layer. This could be either due to a preferred crystallographic direction from the growth on top of the FeF$_2$ or due to short-range order correlations from the FeF$_2$ [22].

It is also worth noting, that the hysteresis curves in Fig. 1 show an asymmetrical shape. One finds, e.g. for $\mu_0 H_{FC} = 0.05$ T first a very sharp edge (when reducing the field) followed by a rounded shape until the negative saturation is achieved. This interesting and unusual behavior is attributed to the strong coupling of the FM to the AF and will be discussed elsewhere [23].

The idea of two oppositely oriented AF subsystems is confirmed by another set of experiments, where the subsystems are reversed by a field step. Figure 2 shows $M(H)$ curves at $T = 10$ K after two different FC procedures: (i) simple FC from 150 to 10 K in $\mu_0 H_{FC1} = 0.05$ T (single negatively shifted loop) and in 0.2 T (single positively shifted loop), or (ii) first FC from 150 to 10 K in $\mu_0 H_{FC1} = 0.05$ T (1), followed by field heating (FH) from 10 K to $T_S$ in 0.05 T (2), then change the field to 0.2 T at $T_S$ (3) and finally FC form $T_S$ to 10 K in $\mu_0 H_{FC2} = 0.2$ T (4). The data obtained from the FC protocol (i) is shown as solid symbols. The curves measured after the FC protocol (ii) are shown with open symbols. In the case of $T_S = 81$ K no effect of the field change is observable. However, with a slightly higher $T_S = 82$ K a double loop is found resembling those shown in Fig. 1. Finally for $T_S = 83$ K virtually no double loop is visible and only a single positively shifted loop is encountered. Interestingly there remains a small signature of the negatively biased loop even until $T_S = 120$ K (data not shown). Two main observations are made: First, it is possible to reverse the biasing direction and even create the 'bi-domain' state upon application of the field step of $\Delta H = H_{FC2} - H_{FC1}$ above a certain temperature. Hence it is possible to nucleate and move the boundaries of the anticipated AF subsystems. The second surprising observation is that the AF 'bi-domain' structure remains stable even above the Néel temperature of $T_N = 78$ K. Although no long range order is present at $T > T_N$ the 'bi-domain' information is stable as evidenced from the reversal experiments. This could be a consequence of a strain-induced enhancement of the AF/FM exchange coupling [24] and therefore a stabilization of the AF by the FM.

In order to investigate the criterion for the occurrence of double hysteresis loops quantitatively, micromagnetic simulations including the influence of the AF are performed. As mentioned above one can expect double hysteresis loops, when the FM does not experience an average AF moment $\langle S_{AF} \rangle$, but if the AF subsystem size is much larger than the FM domain size. In our numerical studies a polycrystalline Ni layer of 20 nm thickness and lateral size of 500 nm × 500 nm is simulated using the OOMMF micromagnetic simulation package [25]. The FM is discretized in three dimensions in cubic cells of size 5 nm. On the first layer ($z=0$) acts an unidirectional, constant and random site field with coverage of about 10% [23]. The field strength corresponds to $J_{AF-FM} = 2 J_{AF}$, where $J_{AF} = -0.45$ meV is the main exchange constant in bulk $FeF_2$ [26]. Apart from standard micromagnetic parameters for Ni we also include the demagnetizing effect by an anisotropy constant, $K_d = -500$ kJ/m$^3$, forcing the spins to be oriented in-plane and an uniaxial anisotropy constant, $K_u = -15$ kJ/m$^3$, which yields an in-plane uniaxial FM axis as found from experiment [23]. We can introduce AF subsystems with opposite field directions with different subsystem sizes, while the total simulated system size is constant. Fig. 3 shows the calculated hysteresis loops, $M(H)$, for different AF subsystem sizes, $D_{AF} = 30, 60, 125,$ and 250 nm.

One clearly observes a transition from one broad loop ($D_{AF} = 30$ nm) where the Ni layer experiences an average moment from the AF to two separated subloops ($D_{AF} = 250$ nm). The latter result is similar to that observed in the experiment. In this case, the sample breaks into two virtually decoupled FM/AF systems. The limiting value necessary to observe double loops has to be compared to the domain size of Ni or better to its domain wall width. This value reflects the length scale on which the FM averages over the AF. For single-crystalline bulk Ni it is: $\delta_B = 82$ nm [27]. In our case we have to include the uniaxial anisotropy term, $K_u = -15$ kJ/m$^3$, thus arriving at $\delta_B = 48$ nm. The criterion, $D_{AF} >> \delta_B$, [15] is well fulfilled in the case studied here. However, more advanced studies should follow, e.g. including the effect of rotatable AF moments and the possibility of a depth dependent magnetization structure in the AF (planar domain walls).

In conclusion, we present an EB system, that exhibits a double hysteresis loop and hence two oppositely biased subsystems ('bi-domain' state). Each 'domain' acts as a local bias field on the FM. This case is found for intermediate cooling fields, $0.075$ T $\leq \mu_0 H_{FC} \leq 0.125$ T. A reversal or movement of the AF 'domains' can be induced by a field step through thermal activation. The essential criterion for obtaining independent subsystems is the size of the AF subsystems, that has to be much larger than the domain size (or domain wall width) of the FM. This idea is confirmed by micromagnetic simulations of a Ni layer with the effect of the AF modeled by a constant, unidirectional, random-site field acting on the FM surface.

We thank for fruitful discussions with W. Kleemann and X. Chen. The work is funded by the US Department of Energy and the AFOSR. Also, financial support from Alexander-von-Humboldt Foundation (O. P.), Cal(IT)$^2$ (Z.-P. L.), Spanish MECD (R. M., X. B.), Fulbright Commission (R. M.), and Catalan DURSI (X. B.) is acknowledged.

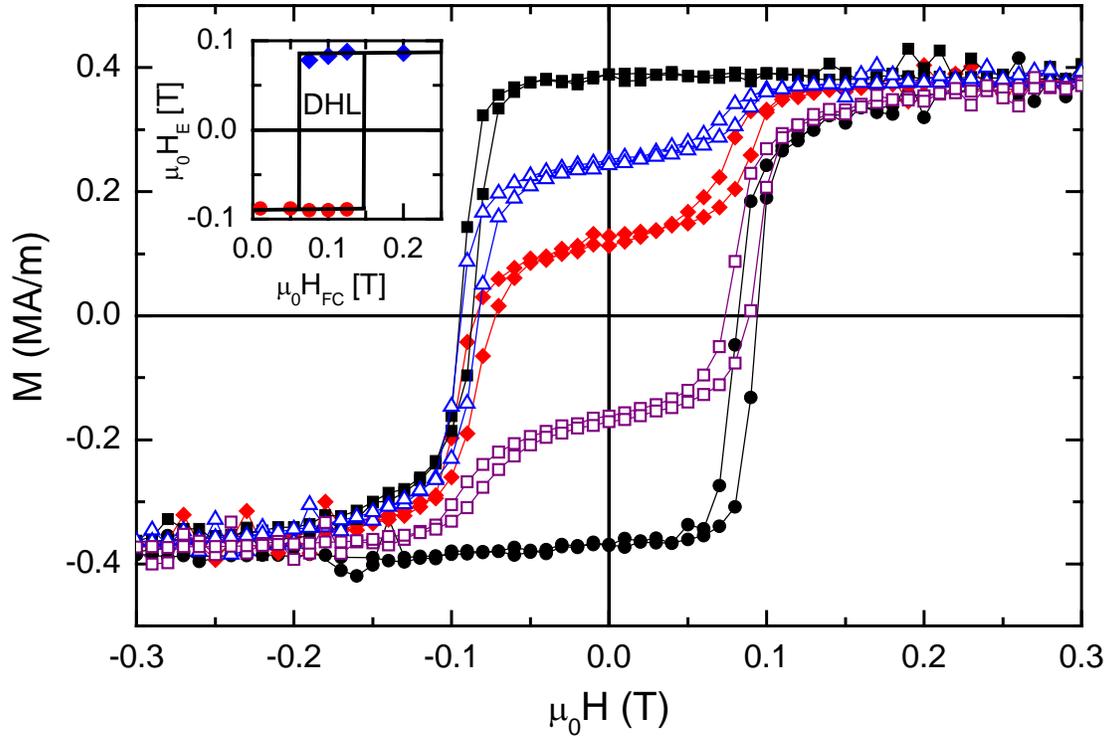

Fig. 1. $M(H)$ at $T = 10$ K after field cooling in $\mu_0 H_{FC} = 0.05$ T (solid squares), 0.075 T (open triangles), 0.1 T (solid diamonds), 0.125 T (open squares) and 0.2 T (solid circles). The inset shows a plot of the extracted EB field $H_E$ vs. FC field $H_{FC}$. The field range, where a double hysteresis loop (DHL) occurs is presented as a rectangle. Solid lines are only guides to the eye.

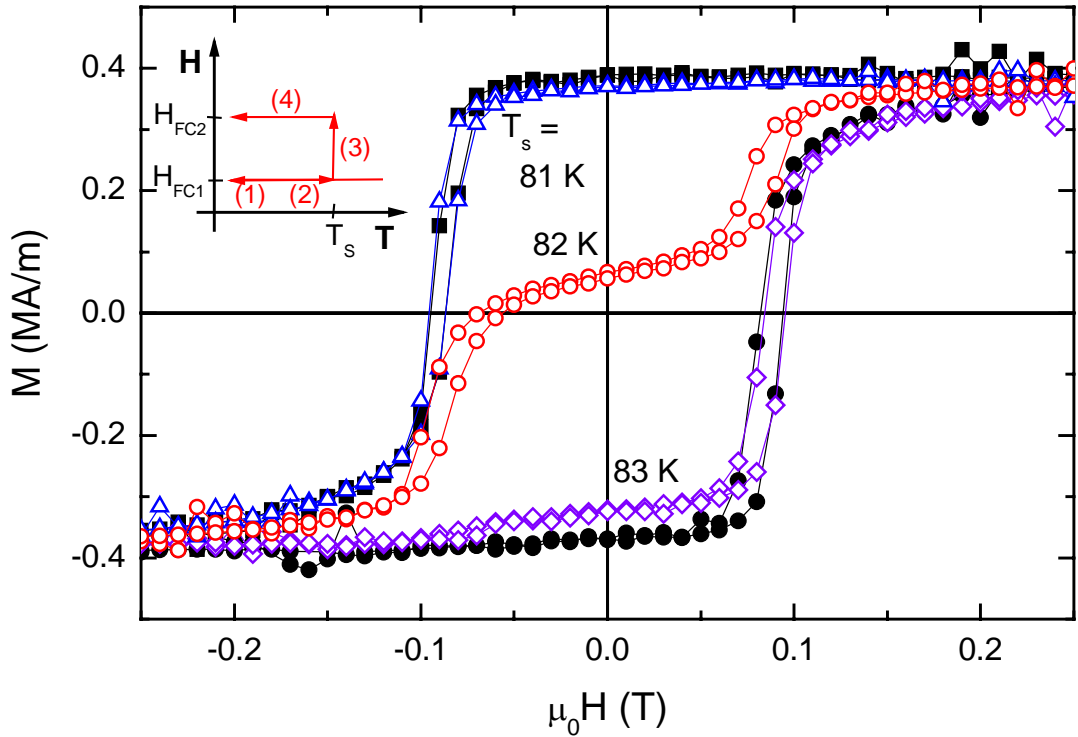

Fig. 2. $M(H)$ at $T = 10$ K for two different FC protocols. (i) Simple FC in $\mu_0 H_{FC} = 0.05$ T (solid squares) and 0.2 T (solid circles) and (ii) after a field step (see text) at $T_S = 81$ K (open triangles), 82 K (open circles) and 83 K (open diamonds), where $\mu_0 H_{FC1} = 0.05$ T and $\mu_0 H_{FC2} = 0.2$ T. Lines are guides to the eyes. The inset shows a schematic representation of the FC procedure (ii).

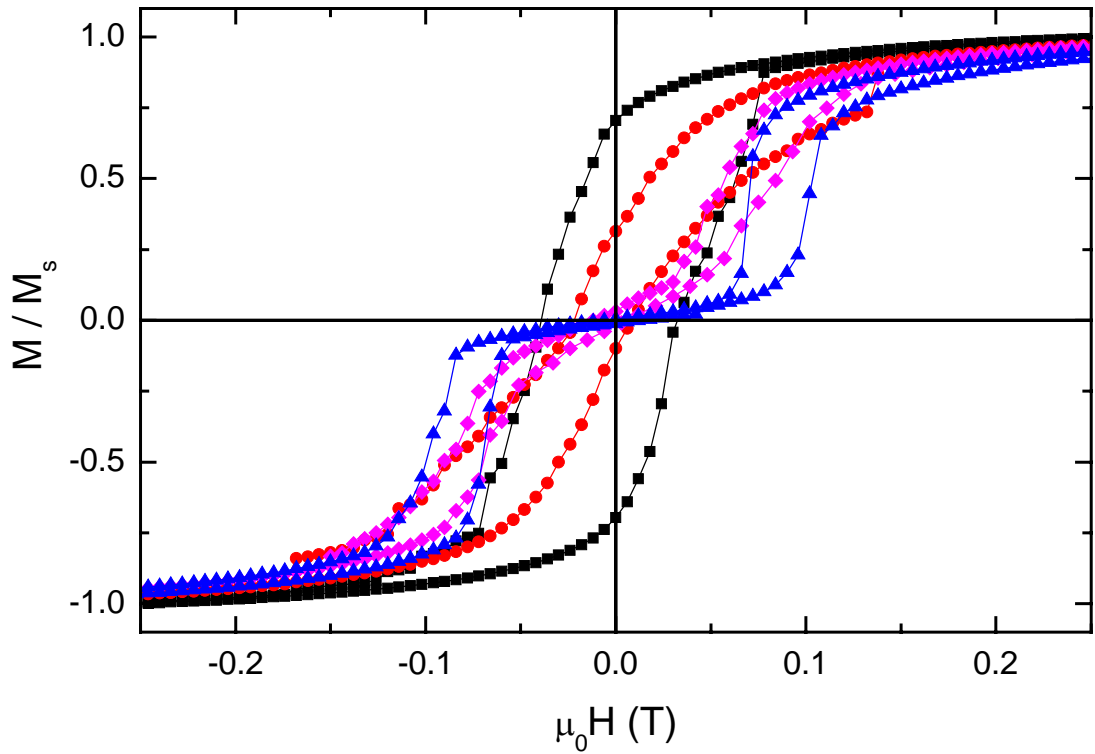

Fig. 3. $M(H)$ from simulations of a Ni layer, where a constant random-site field acts on the bottom layer. The random-site field has two orientations depending on the subsystem. The subsystem size was varied, $D_{AF}=$ 30 (squares), 60 (circles), 125 (diamonds), and 250 nm (triangles). Lines are guides to the eyes.